# Phonon Coherent Resonance and Its Effect on Thermal Transport In Core-Shell Nanowires

Jie Chen,[1] Gang Zhang,[2,*] and Baowen Li[1,3,†]

[1]Department of Physics and Centre for Computational Science and Engineering, National University of Singapore, Singapore 117542, Singapore

[2] Key Laboratory for the Physics and Chemistry of Nanodevices and Department of Electronics, Peking University, Beijing 100871, People's Republic of China

[3]NUS Graduate School for Integrative Sciences and Engineering, Singapore 117456, Singapore

## Abstract

We study heat current autocorrelation function and thermal conductivity in core-shell nanowires by using molecular dynamics simulations. Interestingly, a remarkable oscillation effect in heat current autocorrelation function is observed in core-shell NWs, while the same effect is absent in pure silicon nanowires, nanotube structures and random doped nanowires. Detailed characterizations of the oscillation signal reveal that this intriguing oscillation is caused by the coherent resonance effect of the transverse and longitudinal phonon modes. This phonon resonance results in the localization of the longitudinal modes, which leads to the reduction of thermal conductivity in core-shell nanowires. Our study reveals that a coherent mechanism can be used to tune thermal conductivity in core-shell nanowires.

---

*zhanggang@pku.edu.cn; †phylibw@nus.edu.sg

# 1. Introduction

Thermoelectric materials with low thermal conductivity are favorable for high-efficiency power generation and refrigeration. In the past few years, some significant progresses have been achieved [1, 2] in enhancing thermoelectric performance of silicon, which is abundant in nature, well-engineered in semiconductor industry and friendly to the environment [3]. For example, by etching the surface of silicon nanowires (SiNWs), it has been demonstrated that thermal conductivity of bulk silicon can be reduced more than two orders of magnitude with slight effect on electric power factor, resulting in a dramatically enhanced figure of merit ($ZT$) at room temperature [1].

However, most of the conventional approaches to reduce thermal conductivity, such as introduction of rough surface [1, 2, 4] and impurity [5] scatterings, are based mainly on incoherent mechanisms, which cause phonons to lose coherence, this in turn also deteriorates the electronic transport properties [1]. Recent experimental works [6, 7] have demonstrated that, by altering phonon band structure in periodic nanomesh structures, a remarkable enhancement in $ZT$ can be obtained by significantly reducing thermal conductivity of silicon while preserving its electrical conductivity. These studies have offered new perspective to improve $ZT$ based on coherent mechanisms.

In this paper, by using molecular dynamics simulations, we demonstrate an intriguing phonon coherent resonance phenomenon in Ge/Si core-shell NWs, which offers a coherent mechanism to tune thermal conductivity in core-shell NWs. As Ge/Si core-shell NWs can be synthesized experimentally [8-10], our study suggests novel insights into the thermal management at nanoscale.

## 2. Core/Shell Nanowire Structure and the Coherent Resonance

The configuration of [100] Ge/Si core-shell NWs is shown in Fig. 1. The cross sections of both core and shell regions are square, with side length $L_c$ and $L$, respectively. The longitudinal direction is set along $x$ axis, with $N_X$, $N_Y$, $N_Z$ unit cells (8 atoms per unit cell) in $x$, $y$ and $z$ direction, respectively. Stillinger–Weber (SW) potential [11, 12] is used in our simulations to derive the force term. The force field parameters for Si-Si bond and Ge-Ge bond are from Ref. 11 and Ref. 12, respectively. For Si-Ge bond, the net length and energy units in SW potential are taken to be the arithmetic average and geometric average of that of Si and Ge atom, respectively [13]. The velocity Verlet algorithm is employed to integrate Newton's equations of motion numerically, and time step is set as 0.8 fs.

The coherence of phonons can be probed by the heat current autocorrelation function (HCACF) in equilibrium molecular dynamics (EMD) simulations. The canonical ensemble molecular dynamics simulations with Langevin heat reservoir first runs for $10^5$ steps to equilibrate the heterostructure NWs at a given temperature, during which the free boundary condition is applied to all the atoms on the surface. After structure relaxation, all atoms are assigned with a random velocity according to Gaussian distribution. The periodic boundary condition is applied in the $x$ (longitudinal) direction, and the free boundary condition is applied in the other two directions. Then microcanonical ensemble EMD runs for $3\times10^6$ time steps (2.4 ns) and heat current [14] is recorded at each step. Finally, the heat current (in $x$ direction) autocorrelation function is calculated [15]. The whole procedure is repeated six times with different initial conditions of the velocity distribution.

Fig. 2a shows the typical time dependence of normalized HCACF in a 16×5×5 super cell at 300 K for Ge/Si core-shell NWs with a given core-shell ratio ($L_c/L$=0.65).

For comparison, the HCACF for SiNWs and silicon nanotubes (SiNTs) [4] are also shown in Fig. 2. For both SiNWs and SiNTs, there is a very rapid decay of HCACF at the beginning, followed by a long-time tail with a much slower decay. This two-stage decaying characteristic of HCACF has been reported in the studies of various single-component materials [16-18]. When time is long enough, the long-time tail of HCACF decays to approximately zero.

However, an obvious oscillation up to a very long time appears in HCACF for core-shell NWs (Fig. 2a). The long-time region of HCACF reveals that this oscillation is not random instead it has a periodic structure (Fig. 2b). This has been further verified by extending the total number of time steps in microcanonical ensemble EMD simulations, from $N=3\times10^6$ to $N=5\times10^6$. We find that EMD simulations with even longer simulation time give almost identical result to that shown in Fig. 2, and the periodic oscillation feature is not affected at all.

By using the same EMD simulation procedure, we have also calculated HCACF for $Si_{1-x}Ge_x$ randomly doped NWs with different doping concentration $x$. As shown in Fig. 2c and 2d, none of the $Si_{1-x}Ge_x$ NW exhibits any oscillation in HCACF, consistent with the theoretical idea that the periodic oscillation is a coherent wave effect that requires long-time correlation, which should not take place in a randomly doped heterostructure.

It is worth pointing out that oscillation in HCACF has been reported by Landry et al. [19] in the study of superlattice structures. They found that the oscillation in their work is caused by specific zero-wave-vector optical phonon modes and can be removed when a different definition of the heat current based on the equilibrium atom positions is used [19].

In this work, we have also calculated the heat current according to the definition based on equilibrium atom positions. In contrast, we found that the oscillation in HCACF remains. Furthermore, to verify the generality of this oscillation feature of HCACF in core-shell structure, we have also considered the following cases: rectangular Si/Ge core-shell NWs, circular Ge/Si core-shell NWs (both core and shell regions are circular), and rectangular Ge/Si core-shell with imperfect interface (e.g., 5% and 10% of the atoms at the interface are randomly switched). As shown in Fig. 3, oscillations in HCACF still exist in these cases, suggesting that it is a generic characteristic in core-shell structures. For the sake of simplicity for theoretical modeling and analysis, all the Ge/Si core-shell NWs considered in the rest of this study are rectangular cross sections as depicted in Fig. 1.

Fig. 4a and 4b shows the long-time region of normalized HCACF for Ge/Si core-shell NWs with different core-shell ratio at 300 K. This oscillation effect exhibits an obvious structure dependence: when the core-shell ratio increases, it becomes stronger, reaches its maximum amplitude at $L_c/L=0.65$, and then decreases. Moreover, for a given core-shell structure, the oscillation amplitude is temperature-dependent and becomes larger at lower temperature (Fig. 4c). The structure and temperature dependence of the oscillation in HCACF suggests that there exists a coherent mechanism in core-shell NWs which can cause phonons to have the long-lasting correlation in such heterostructure.

To quantitatively characterize the oscillation effect in core-shell NW, we numerically measure the oscillation amplitude. Consider a perfect cosine oscillation function described by $u = A\cos(\omega t + \varphi)$, the standard deviation ($\sigma$) of this cosine function in the time interval ($t_1$, $t_1+\Delta t$) is given by $\sigma^2 = \frac{1}{\Delta t}\int_{t_1}^{t_1+\Delta t} A^2 \cos^2(\omega t + \varphi)dt = \frac{A^2}{2}$, provided that $\Delta t$ is a multiple of the period. So the amplitude of a periodic oscillation can be measured by calculating the standard deviation ($A=\sqrt{2}\,\sigma$). For the random noise, the standard deviation is indeed a good quantity to gauge the noise level. Here we calculate the standard deviation of the long-time region of HCACF from $t_d$ to $t_d+\Delta t$, where $t_d$ is the time after HCACF (or the envelope of HCACF for oscillation case) decays to approximately zero (e.g., $t_d$=15 ps in Fig. 2b), and $\Delta t$ is set as 15 ps during which the oscillation amplitude is almost constant.

For each core-shell structure, the final results are averaged over six measurements of different realizations of HCACF. For example, for those realizations of HCACF shown in Fig. 2b, the standard deviation of HCACF for Ge/Si core-shell NWs calculated from 15 ps to 30 ps is *σ=0.031*, and the calculated amplitude is *A=$\sqrt{2}\,\sigma$=0.044*, consistent with the oscillation amplitude shown in Fig. 2b. For SiNWs, the calculated standard deviation is about *σ=0.003*, and its corresponding amplitude is *A=0.0042*, which gives an estimation of the noise level and is one order of magnitude smaller than the amplitude of the oscillation in Ge/Si core-shell NWs. This method to measure oscillation amplitude can give distinct estimations of the nontrivial oscillation and the noise level. Fig. 4d shows the results of the structure- and temperature-dependent oscillation amplitude. With core-shell ratio increases, the oscillation amplitude first increases, reaches a peak value at $L_c/L=0.65$ and then decreases. More interestingly, the oscillation amplitude at different temperature shows the same structure dependence, with larger amplitude at low temperature and vanishing amplitude (comparable to noise level) at high temperature.

## 3. Physical Mechanism of the Coherent Resonance

To better understand the underlying mechanism, we have carried out extensive spectrum analysis by using the fast Fourier transform (FFT). We calculate the FFT of the long-time region of normalized HCACF from time $t_d$ to $t_d+\Delta t$, with $\Delta t$ set as 15 ps. It has been further checked that our FFT analysis is robust with the particular choice of $t_d$. Fig. 5 shows the FFT of normalized HCACF for different structures shown in Fig. 2b. The FFT amplitude for Ge/Si core-shell NWs exhibits a dominant peak at low frequency, which is denoted as $f_0$ in Fig. 5a. All the FFT amplitudes shown in Fig. 5 are normalized by the amplitude of the dominant peak $f_0$. Moreover, there exist multiple high frequency peaks shown in Fig. 5b, with much smaller amplitude compared to that of the dominant peak. The FFT spectrum of SiNWs (Fig. 5c) and SiNTs (Fig. 5d) looks completely different from that of Ge/Si core-shell NWs: there is no dominant peak over the entire frequency regime, and the FFT amplitude is more than 2 orders of magnitude smaller than the dominant peak amplitude for Ge/Si core-shell NWs. These two aspects are the typical characteristics of the noise spectrum. This spectrum analysis verifies that there is no significant oscillation in the long-time region of HCACF for SiNWs and SiNTs, and the fluctuation of HCACF in NW/NT is mainly due to the computational noise.

The multiple oscillation peaks observed in frequency domain for core-shell NWs are very similar to the confinement effect of the acoustic wave (coherent long wavelength phonon) in a confined structure. For a wire with the square cross section and side length $L$, the eigen-frequency of the transverse modes in such confined structure calculated from the elastic medium theory is given by:

$$f_{mn} = \frac{\omega_{mn}}{2\pi} = \frac{k_{mn}C}{2\pi} = \frac{C}{2\pi}\sqrt{\left(\frac{m\pi}{L}\right)^2 + \left(\frac{n\pi}{L}\right)^2} = \frac{C}{2L}\sqrt{m^2+n^2} = \sqrt{m^2+n^2}f_0, \quad (1)$$

where $f_{mn}$ is the eigen-frequency specified by two integer number $m$ and $n$, $C$ is the speed of sound, and $f_0=C/2L$ is the lowest eigen-frequency. We record the high frequency peaks marked by the black arrows in Fig. 5b, and compare them with the frequency of the dominant peak $f_0$ in Fig. 5a. We find the relation between the frequencies of these high frequency peaks and the dominant peak $f_0$ is very close to that given by Eq. (1). This good agreement of oscillation frequency suggests that the intriguing oscillation effect results from the frequency quantization of the transverse modes as a consequence of structure confinement in the transverse direction.

In single-component homogeneous NWs, atoms on the same cross section plane have the same sound velocity, so that the transverse motion is decoupled from the longitudinal motion. This is why the oscillation signal is not probed by the longitudinal HCACF in SiNWs. In core-shell NWs, atoms on the same cross section plane have different sound velocity in the longitudinal direction. As a result, atoms near the core-shell interface are stretched due to the different sound velocity. This induces a strong coupling/interaction between the transverse and longitudinal motions.

It is well known that when there is an interaction between two modes, the oscillation amplitude will be maximized when the frequencies of these two modes are close to each other (resonance). Due to the frequency quantization of the transverse modes, resonance will take place when the frequency of the longitudinal mode is close to the eigen-frequency of the transverse mode, giving rise to the enhanced oscillation amplitude. This coupling picture explains that frequency quantization of the transverse modes can indeed manifest itself in HCACF along the longitudinal direction in core-shell NWs, while the same effect is absent in SiNWs, NTs and random doped NWs. Moreover, as the resonance effect of acoustic wave is a coherent process that requires long-time correlation, the stronger anharmonic phonon-phonon scattering at high temperature causes phonon to lose coherence, and leads to the vanishing of the oscillation effect at high temperature.

To quantitatively characterize the coupling between the transverse and longitudinal modes in the core-shell structure, we consider the following two scenarios. For a core-shell structure consists of two homogeneous mediums A (core) and B (shell), if there is no coupling between A and B at the interface, the speed of sound $C$ in this inhomogeneous medium (A/B core-shell) is either $C_{core}=C_A$ in medium A, or $C_{shell}=C_B$ in medium B, depending on the position. This is the scenario without the interface coupling. However, due to the coupling at the interface between A and B, the speed of sound in this inhomogeneous medium is different from that of its pure element, and can be effectively described by another homogeneous medium E with the same elastic properties. This is the spirit of the effective medium approximation (EMA) and the actual case (with coupling). Since the mode coupling in core-shell structure is induced by the mismatch of sound velocity, we propose to define the coupling strength $S$ in core-shell structure to be the difference between these two scenarios (with and without coupling) as:

$$S = \frac{N_{core}(C_{core}-C_{eff})^2 + N_{shell}(C_{shell}-C_{eff})^2}{N}, \quad (2)$$

where $N$ is the total number of atoms in core-shell structure, $N_{core}/N_{shell}$ is the number of atoms in the core/shell region, $C_{eff}$ is the effective speed of sound in core-shell structure, and $C_{core}/C_{shell}$ denotes the speed of sound in pure core/shell medium. According to this definition, the coupling strength is zero for pure SiNWs, consistent with the result that there is no resonance effect in SiNWs.

The speed of sound in pure Si and Ge NWs, and effective speed of sound in the Ge/Si core-shell NWs are calculated by using "General Utility Lattice Program" (GULP) package [20]. Bulk modulus $K$, Shear modulus $G$ and mass density $\rho$ of each structure are calculated by GULP after structure relaxation. Then the speed of sound is calculated according to $C = \sqrt{\left(K + \frac{4}{3}G\right)/\rho}$. As shown in Fig. 6, the dependence of coupling strength on core-shell ratio agrees qualitatively well with the variation of the measured oscillation amplitude shown in Fig. 4d. This good agreement reveals that the structure dependence of the oscillation amplitude is caused by the structure-dependent coupling strength.

## 4. Coherent Resonance Induced Phonon Localization and Reduction of Thermal Conductivity

Due to the nonpropagating nature of the transverse modes, the resonance effect induced by coupling between the transverse and longitudinal modes can hinder the longitudinal transport [21], thus offering a coherent mechanism to tune thermal conductivity in core-shell structure. To illustrate this coherent mechanism in more details, we study the localization effect of the longitudinal phonon modes in SiNWs and Ge/Si core-shell NWs. All phonon modes are computed by using GULP with a cross section of 5×5 unit cells. Mode localization can be quantitatively characterized by the phonon participation ratio $P_\lambda$ defined for each eigen-mode $\lambda$ as [22]

$$P_\lambda^{-1} = N \sum_i \left( \sum_\alpha \varepsilon_{i\alpha,\lambda}^* \varepsilon_{i\alpha,\lambda} \right)^2, \qquad (3)$$

where $N$ is the total number of atoms, and $\varepsilon_{i\alpha,\lambda}$ is the $\alpha$-th eigenvector component of eigen-mode $\lambda$ for the $i$-th atom. When considering the overall influence of structure change on localization effect, averaged phonon participation ratio ($P_{ave}$) defined as:

$$P_{ave} = \sum_\omega P(\omega) DOS(\omega), \qquad (4)$$

where $DOS(\omega)$ is the density of states. $P_{ave}$ is a very useful quantity to characterize the overall localization effect, with a smaller $P_{ave}$ indicating a stronger phonon localization effect [13]. Compared to SiNWs, $P_{ave}$ in Ge/Si core-shell NWs shows an overall reduction (Fig. 7a). More interestingly, $P_{ave}$ in Ge/Si core-shell NWs shows a non-monotonic variation with the increase of core-shell ratio. This is difficult to be understood from the conventional diffusive phonon picture, which suggests that the interface scattering becomes stronger with the increase of interface [23]. Our coupling picture can give a qualitatively understanding of such non-monotonic variation: the resonance effect induced by coupling between the transverse and longitudinal phonon modes causes localization effect of the longitudinal phonon modes, and the strength of the localization effect $P_{ave}$ is controlled by the coupling strength $S$, with a smaller $P_{ave}$ at larger $S$.

It is reported in the previous study of single polyethylene chains that there exists non-periodic and non-attenuating oscillation in the normal mode autocorrelation function [24], which is considered to be responsible for the unusually high thermal conductivity in the individual polymer chain [24, 25]. In this work, however, the oscillation in HCACF observed in core-shell NW is periodic and it oscillates around zero. Moreover, as explained above, the oscillation in HCACF is induced by the phonon resonance, which can cause phonon localization and hence lead to the reduction of thermal conductivity.

To understand how much the phonon localization can affect the thermal transport, we further investigate thermal conductivity κ of Ge/Si core-shell NWs. We should point out that when applying the Green-Kubo method to calculate thermal conductivity, it is a challenge to accurately specify the converged values of HCACF integral [26]. Because of the oscillation in HCACF observed in our calculations, it is even more challenging to specify the converged value of the HCACF integral for core-shell NWs. Thus we use non-equilibrium molecular dynamics (NEMD) simulations with Langevin heat reservoir to predict the thermal conductivity. In NEMD calculation of thermal conductivity of bulk material, the calculation results will depend on the size of the simulation cell if there are not enough phonon modes to accurately describe the phonon transport process. In this situation, extrapolation procedure must be applied to NEMD results with multiple system sizes. However, as demonstrated by Sellan *et al.* [26], the commonly used linear extrapolation procedure is not always accurate in Stillinger-Weber silicon. Moreover, in this paper, our concern is the relative reduction of thermal conductivity in core-shell NWs, rather than its absolute value. Thus thermal conductivity of the single-component NWs with the same cross section area and length is used as the reference.

Our NEMD simulations are performed long enough ($4\times10^7$ time steps) to ensure the non-equilibrium steady state. Fig. 7b shows the structure dependent thermal conductivity in Ge/Si core-shell NWs with different cross section areas at 300 K. For all cross section areas, thermal conductivity can be reduced compared to that of SiNWs after introducing the core-shell structure. For the Ge/Si core-shell NWs with cross section of 5×5 unit cells, the trend of the structure-dependent thermal conductivity agrees quite well with that of $P_{ave}$ as shown in Fig. 7a. With larger cross section areas, the structure dependence of thermal conductivity is qualitatively the same, with quantitative difference presumably due to the surface and diameter effect. These results suggest that the phonon localization, which is induced by coherent resonance effect, is responsible for the structure-dependent reduction of thermal conductivity in Ge/Si core-shell NWs. However, as the surface effect is inherently considered in molecular dynamics simulations, the quantitative contribution from coherent resonance effect to the reduction of thermal conductivity in core-shell NWs is still an open question and needs further study.

As demonstrated in Fig. 4d, the stronger anharmonic phonon-phonon scattering at high temperature causes phonon to lose coherence, and leads to the vanishing of the resonance effect at high temperature. To further investigate the effect of coherent resonance on the reduction of thermal conductivity in core-shell NWs, we finally calculate the temperature dependence of thermal conductivity reduction. Thermal conductivity of Ge/Si core-shell NWs with a cross section of 5×5 unit cells and core-shell ratio $L_c/L$=0.65 is calculate and compared with that of GeNWs with the same cross section area at each temperature, taking into account the quantum correction [27, 28]. Fig. 8 shows that the reduction of thermal conductivity is more significant at lower temperature. In addition, our simulation results show that it is possible to further reduce κ of low thermal conductivity material (Ge) by replacing the outer layers with high thermal conductivity material (Si) to form the core-shell structure.

## 5. Conclusions

In summary, we have systematically studied the phonon coherent resonance and its effect on thermal conductivity in core-shell NWs. We found an intriguing oscillation in HCACF in Ge/Si core-shell NWs, which is robust to the change of shape and detailed structure of the interface. This oscillation is absent in pure silicon NWs, nanotube structures and random doped NWs. Detailed characterizations of the oscillation signal uncover that the underlying physical mechanism of this oscillation is the coherent resonance effect of the transverse and longitudinal phonon modes, which is induced by the mode coupling in core-shell NWs. The phonon resonance causes the localization of the longitudinal phonon modes and consequently hinders phonon transport along the core-shell NWs. As a result, thermal conductivity of Ge/Si core-shell NWs can be tuned by the strength of the coherent resonance effect.

Moreover, as the coherence of phonons can be better preserved at lower temperature, it is found that the reduction of thermal conductivity in core-shell NWs is more significant with the decrease of temperature. Our study suggests that the conventional picture of diffusive phonon transport is not sufficient to describe thermal transport in core-shell NWs, and atomistic approaches without assumptions about the nature of phonon transport are suggested. More importantly, our study reveals a coherent mechanism to tune thermal conductivity in core-shell NWs by engineering phonon resonance.

**Acknowledgements**

The authors thank L.-F Zhang for helpful discussions, R. Chen and J. Xiang for sharing us their preliminary experimental results of the core-shell Nanowires and inspiring discussions. This work was supported in part by a grant from the Asian Office of Aerospace R&D of the US Air Force (AOARD-114018), and a grant from SERC of A*STAR (R-144-000-280-305), Singapore, GZ was supported by the Ministry of Science and Technology of China (Grant Nos. 2011CB933001).

**Reference**

1. A. I. Hochbaum, R. Chen, R. D. Delgado, W. Liang, E. C. Garnett, M. Najarian, A. Majumdar, and P. Yang, *Nature* **451,** 163 (2008).

2. A. I. Boukai, Y. Bunimovich, J. Tahir-Kheli, J. -K. Yu, W. A. Goddard III, and J. R. Heath, *Nature* **451,** 168 (2008).

3. C. B. Vining, *Nature* **451,** 132 (2008).

4. J. Chen, G. Zhang, and B. Li, *Nano Lett.* **10,** 3978 (2010).

5. N. Yang, G. Zhang, and B. Li, *Nano Lett.* **8,** 276 (2008); L. Shi, D. Yao, G. Zhang, and B. Li, *Appl. Phys. Lett.* **95**, 063102 (2009).

6. J.-K. Yu, S. Mitrovic, D. Tham, J. Varghese, and J. R. Heath, *Nature Nanotech.* **5,** 718 (2010).

7. J. Tang, H. -T. Wang, D. H. Lee, M. Fardy, Z. Huo, T. P. Russell, and P. Yang, *Nano Lett.* **10,** 4279 (2010).

8. J. Xiang, W. Lu, Y. Hu, Y. Wu, H. Yan, and C. M. Lieber, *Nature* **441,** 489 (2006).

9. J. Xiang, A. Vidan, M. Tinkham, R. M. Westervelt, and C. M. Lieber, *Nature Nanotech.* **1,** 208 (2006).

10. Y. Hu, H. O. H. Churchill, D. J. Reilly, J. Xiang, C. M. Lieber, and C. M. Marcus, *Nature Nanotech.* **2,** 622 (2007).

11. F. H. Stillinger and T. A. Weber, *Phys. Rev. B* **31,** 5262 (1985).

12. K. Ding and H. C. Andersen, *Phys. Rev. B* **34,** 6987 (1986).

13. J. Chen, G. Zhang, and B. Li, *Appl. Phys. Lett.* **95,** 073117 (2009).

14. S. G. Volz and G. Chen, *Appl. Phys. Lett.* **75**, 2056 (1999).

15. P. K. Schelling, S. R. Phillpot, and P. Keblinski, *Phys. Rev. B* **65,** 144306 (2002).

16. J. Che, T. Cagin, W. Deng, and W. A. Goddard, *J. Chem. Phys.* **113,** 6888 (2000).

17. A. J. H. McGaughey, and M. Kaviany, *Int. J. Heat Mass Transfer* **47,** 1783 (2004).

18. J. Chen, G. Zhang, and B. Li, *Phys. Lett. A* **374,** 2392 (2010).

19. E. S. Landry, M. I. Hussein, and A. J. H. McGaughey, *Phys. Rev. B* **77,** 184302 (2008).

20. J. D. Gale, *J. Chem. Soc., Faraday Trans.* **93,** 629 (1997).

21. Z. Liu, X. Zhang, Y. Mao, Y. Y. Zhu, Z. Yang, C. T. Chan, and P. Sheng, *Science* **289,** 1734 (2000).

22. A. Bodapati, P. K. Schelling, S. R. Phillpot, and P. Keblinski, *Phys. Rev. B* **74**, 245207 (2006).

23. R. Yang, G. Chen, and M. S. Dresselhaus, *Nano Lett.* **5**, 1111 (2005).

24. A. Henry and G. Chen, *Phys. Rev. B* **79**, 144305 (2009).

25. A. Henry and G. Chen, *Phys. Rev. Lett.* **101**, 235502 (2008).

26. D. P. Sellan, E. S. Landry, J. E. Turney, A. J. H. McGaughey, and C. H. Amon, *Phys. Rev. B* **81**, 214305 (2010).

27. S. G. Volz and G. Chen, *Phys. Rev. B* **61**, 2651 (2000).

28. J. E. Turney, A. J. H. McGaughey, and C. H. Amon, *Phys. Rev. B* **79**, 224305 (2009).

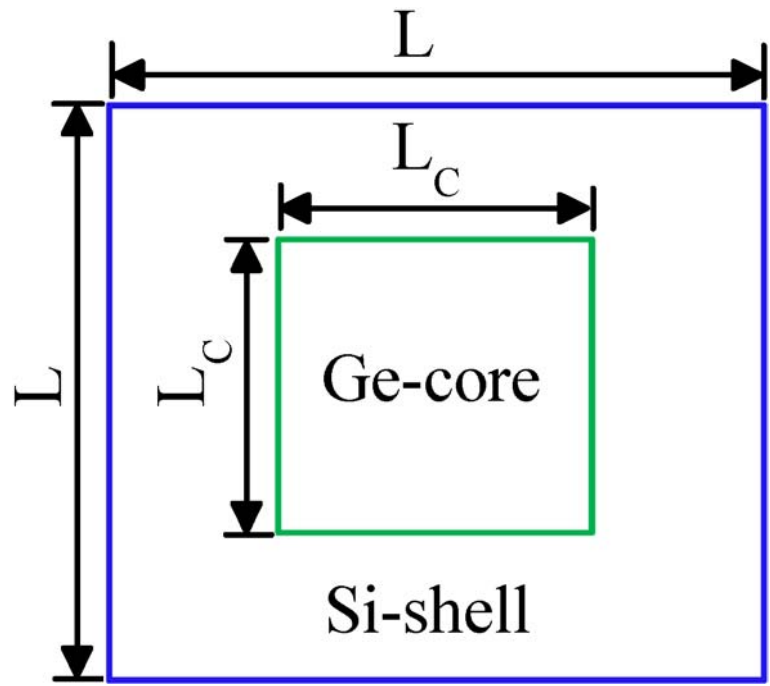

Figure 1. Cross sectional view of [100] Ge/Si core-shell NWs. The cross sections of both core and shell regions are square, with $L_c$ and $L$ denoting the side length of core and shell regions, respectively.

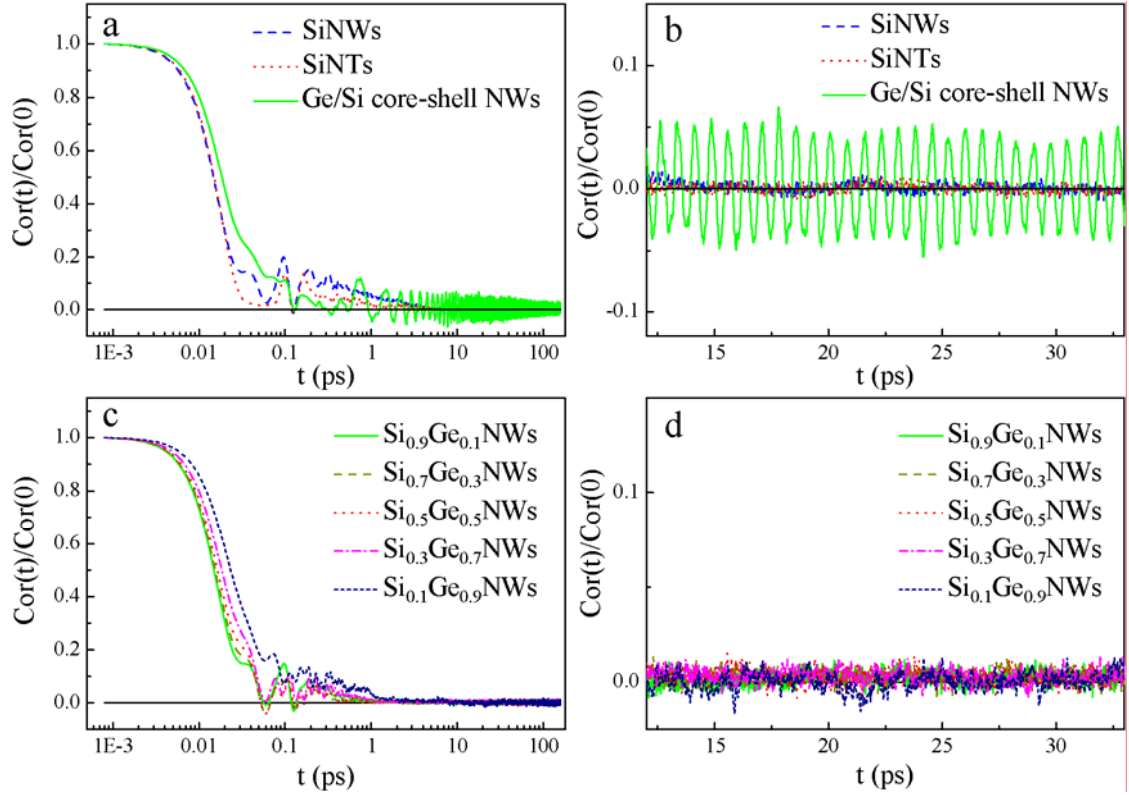

Figure 2. Time dependence of normalized heat current autocorrelation function (HCACF). a, Normalized HCACF for SiNWs (dashed line), SiNTs (dotted line) and Ge/Si core-shell NWs with $L_c/L$=0.65 (solid line). b, Long-time region of a. c, Normalized HCACF for $Si_{1-x}Ge_x$ NWs with different doping concentration $x$. d, Long-time region of c. The black lines in all figures draw the zero axis for reference. Here the super cell size is 16×5×5 and temperature is 300 K.

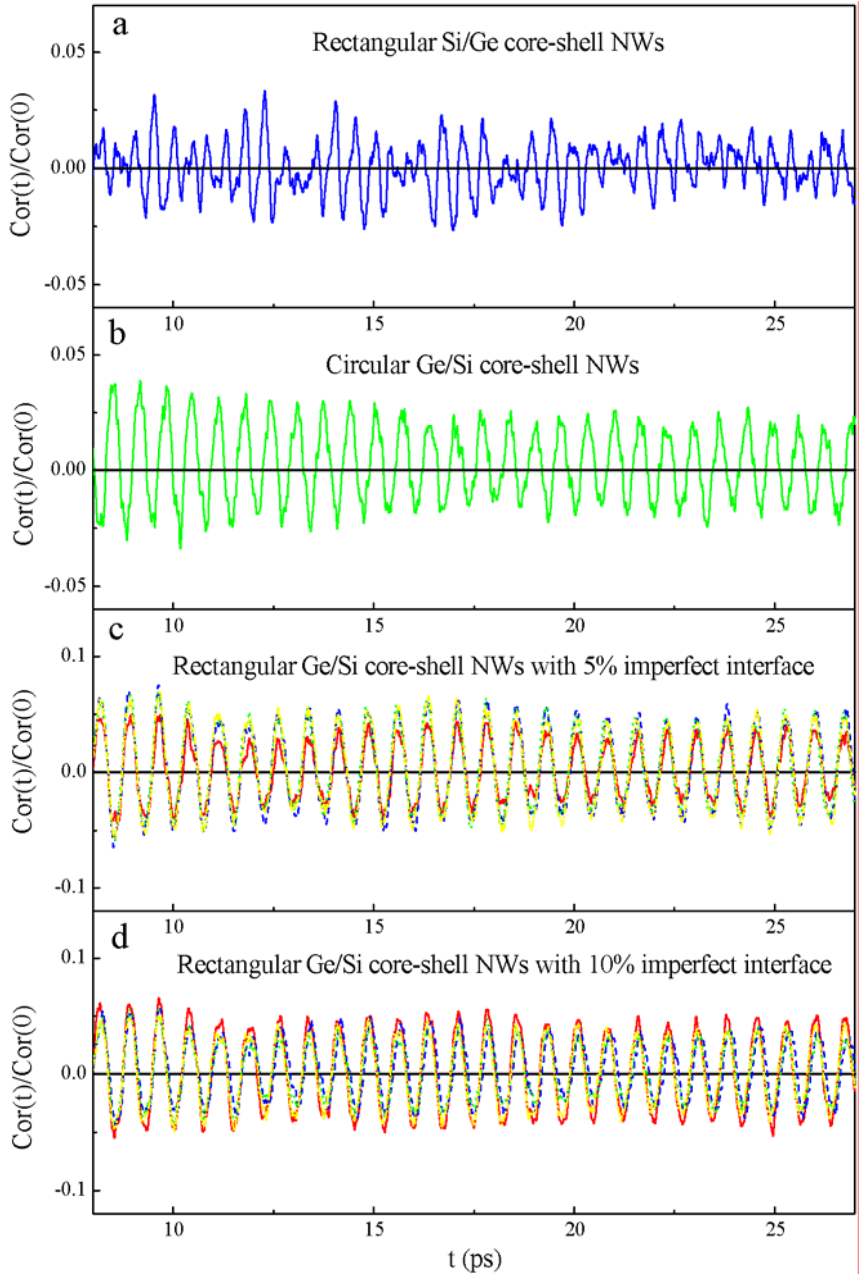

Figure 3. Long-time region of normalized HCACF for different core-shell NWs. a, Rectangular Si/Ge core-shell NWs. b, Circular Ge/Si core-shell NWs. c, Rectangular Ge/Si core-shell NWs with 5% of the atoms at the interface randomly switched. d, Rectangular Ge/Si core-shell NWs with 10% of the atoms at the interface randomly switched. Different colors in c and d denote different realizations for the random switch. The black lines in all figures draw the zero axis for reference.

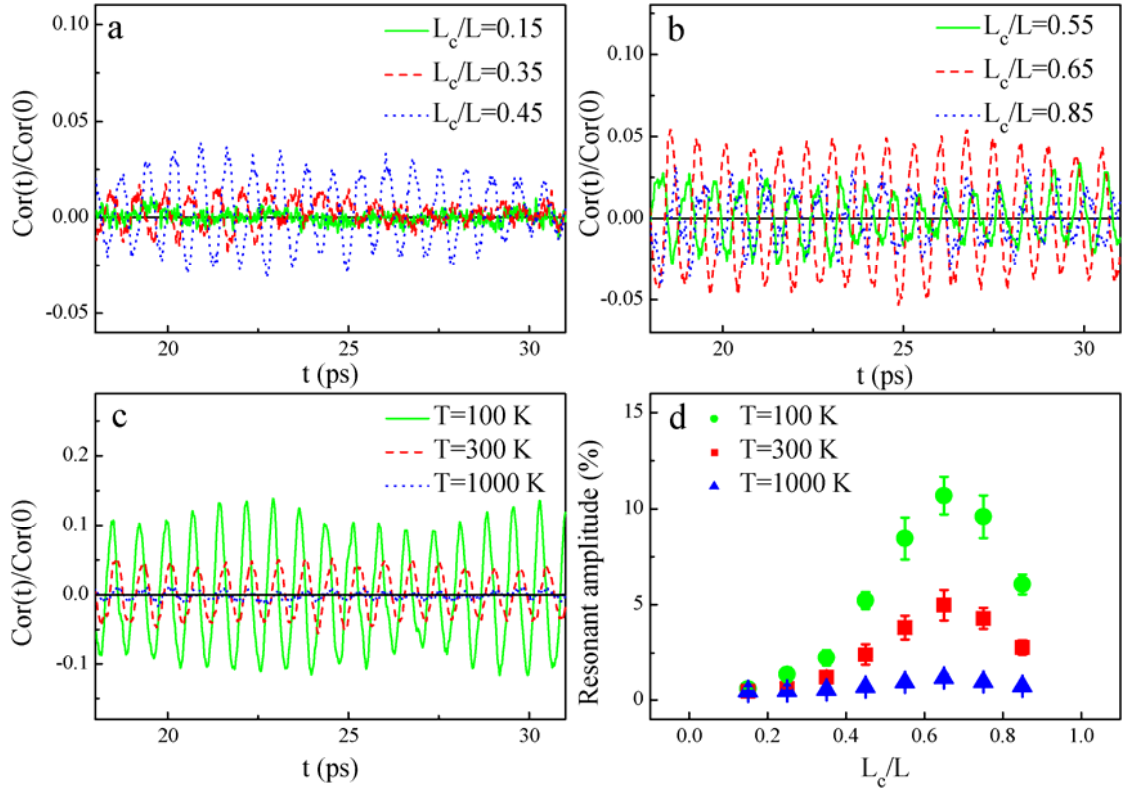

Figure 4. Structure and temperature dependence of the oscillation effect in Ge/Si core-shell NWs. Here we show the long-time region of normalized HCACF in a 16×5×5 super cell. The black lines draw the zero axis for reference. a, $L_c/L$=0.15 (solid line), $L_c/L$=0.35 (dashed line) and $L_c/L$=0.45 (dotted line) at 300 K. b, $L_c/L$=0.55 (solid line), $L_c/L$=0.65 (dashed line) and $L_c/L$=0.85 (dotted line) at 300 K. c, $L_c/L$=0.65 at 100 K (solid line), 300 K (dashed line) and 1000 K (dotted line). d, Oscillation amplitude versus core-shell ratio $L_c/L$ at 100 K (circle), 300 K (square) and 1000 K (triangle).

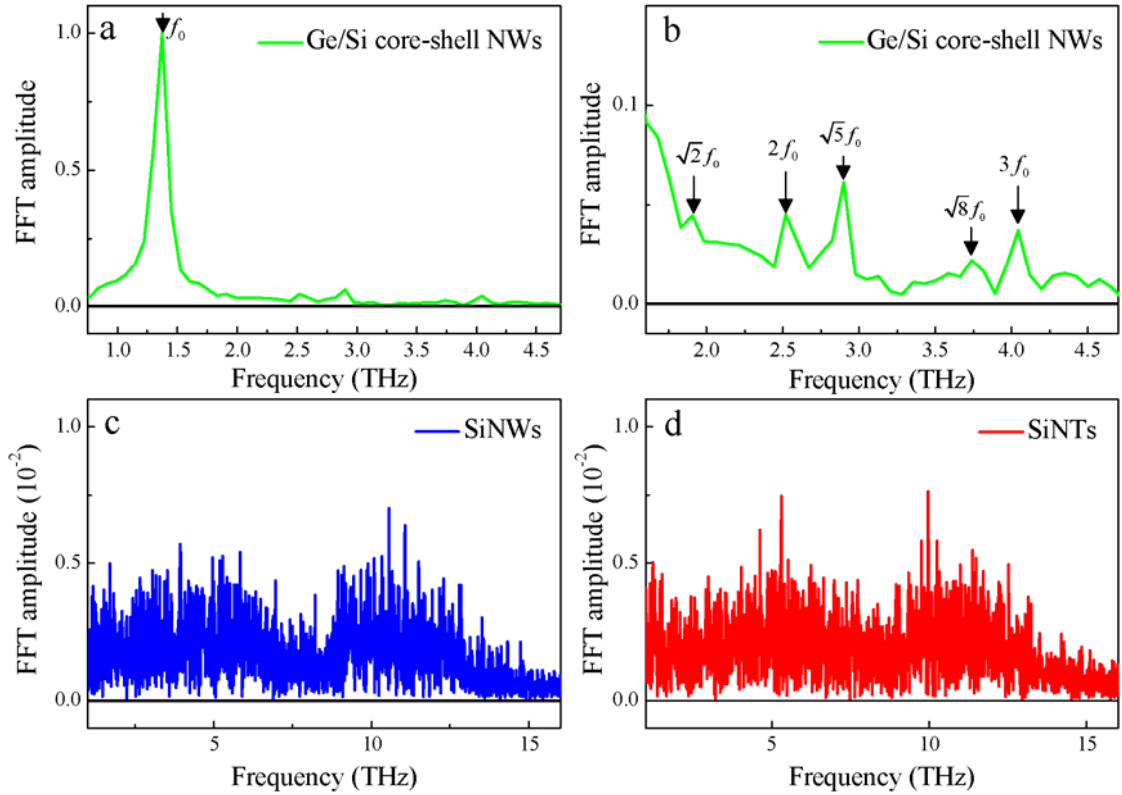

Figure 5. Amplitude of the fast Fourier transform (FFT) of the long-time region of normalized HCACF. The black lines in all figures draw the zero axis for reference. a, Ge/Si core-shell NWs. b, The high frequency oscillation peaks for Ge/Si core-shell NWs. The black arrows pinpoint the different oscillation frequencies. c and d are amplitudes of the FFT of the long-time region of normalized HCACF for SiNWs and SiNTs, respectively.

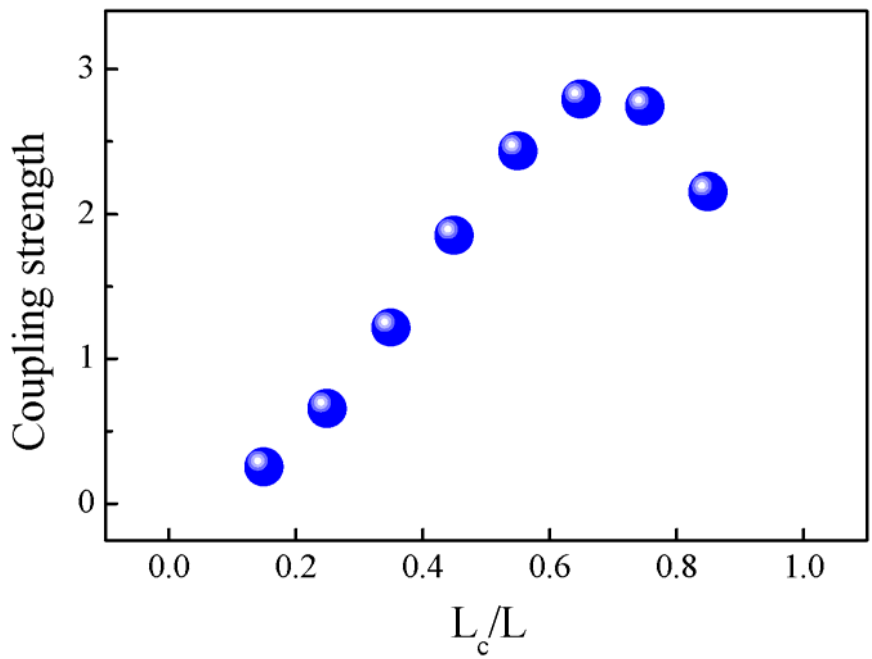

Figure 6. Coupling strength versus core-shell ratio in Ge/Si core-shell NWs.

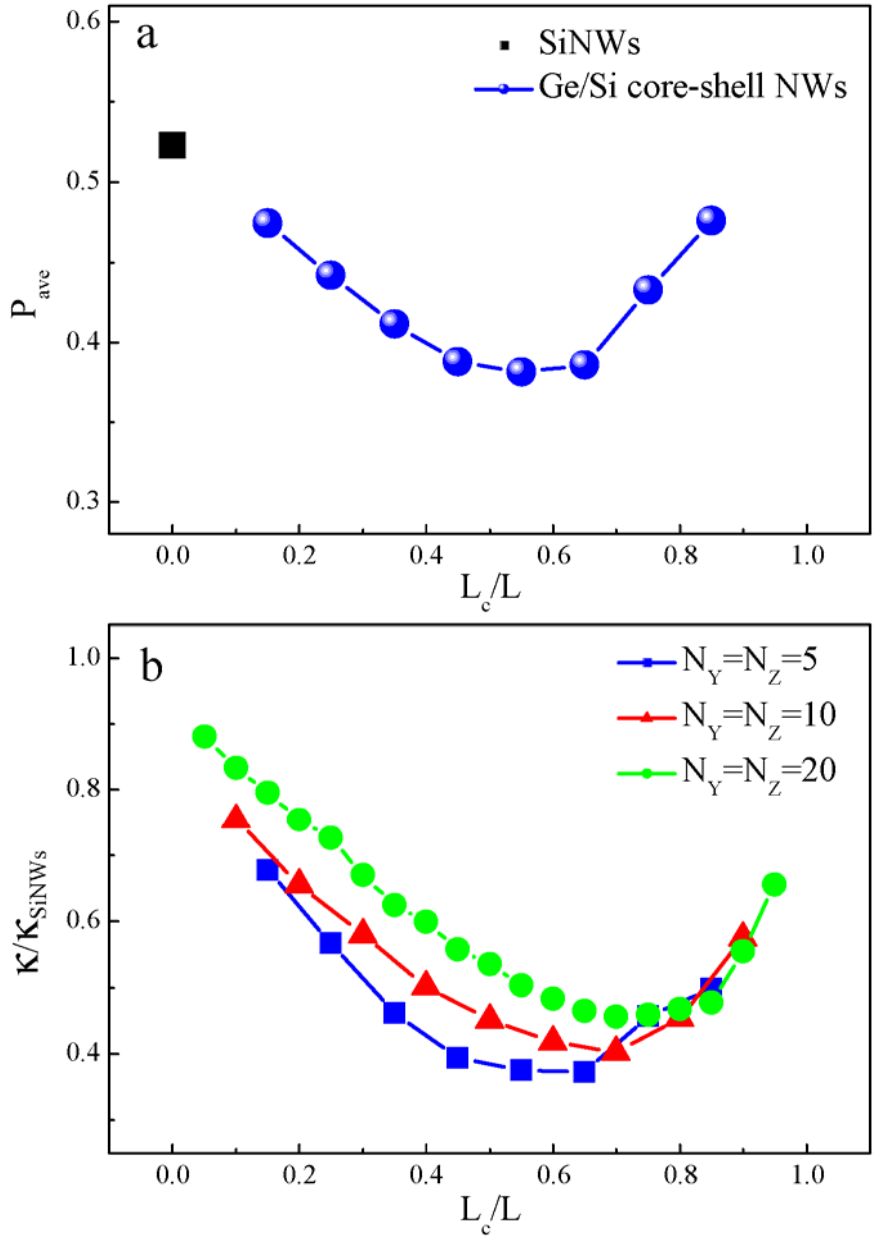

Figure 7. a, Averaged phonon participation ratio $P_{ave}$ versus core-shell ratio in Ge/Si core-shell NWs (circle). The black square plots $P_{ave}$ in SiNWs with the same cross section area for comparison. All phonon modes are computed with a cross section of 5×5 unit cells. b, Structure dependence of thermal conductivity in Ge/Si core-shell NWs with different cross section areas ($N_Y \times N_Z$ unit cells) at 300 K. Thermal conductivity of SiNWs with the same cross section area is used as the reference. The length of the NW is 16 unit cells.

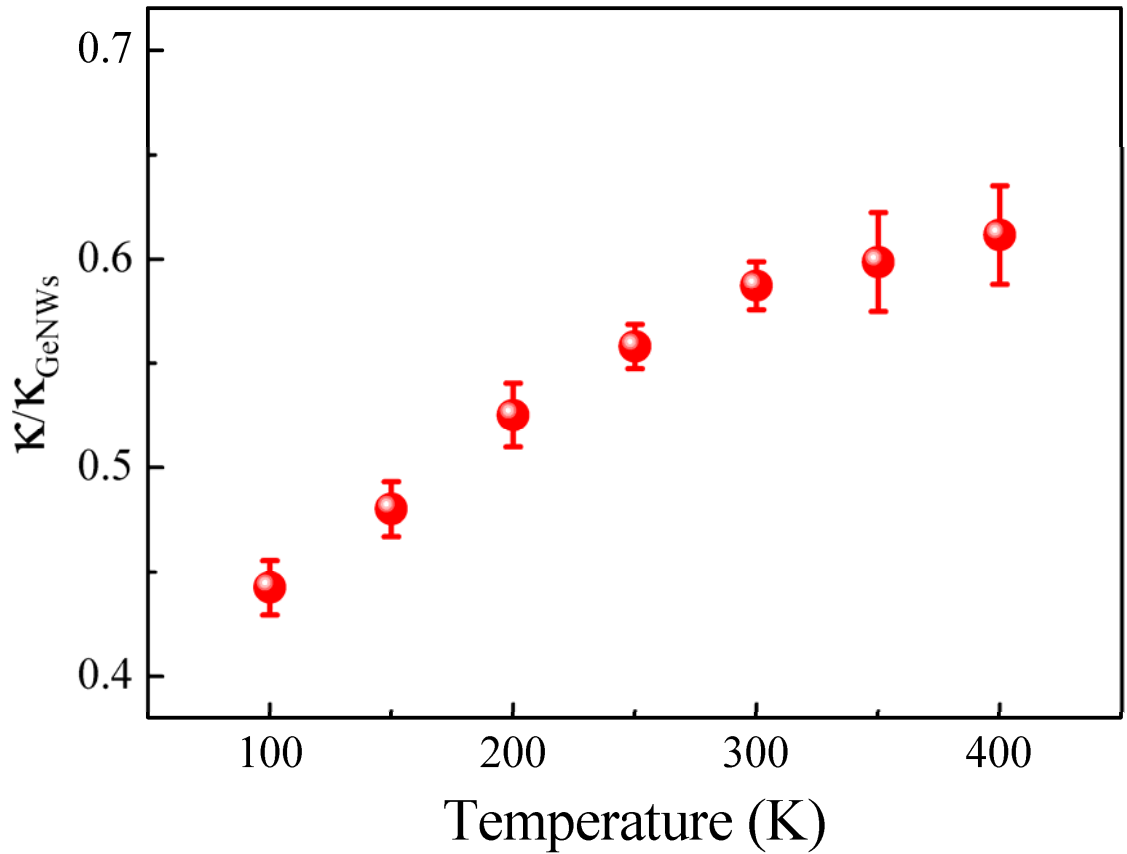

Figure 8. Temperature dependence of thermal conductivity reduction in Ge/Si core-shell NWs. Here thermal conductivity of Ge/Si core-shell NWs with a cross section of 5×5 unit cells and core-shell ratio $L_c/L$=0.65 is compared with that of Ge NWs with the same cross section area at each temperature. The length of the NW is 16 unit cells.